\documentclass[twocolumn,showpacs,preprintnumbers,amsmath,amssymb]{revtex4}

\usepackage{graphicx}% Include figure files
\usepackage{dcolumn}% Align table columns on decimal point
\usepackage{bm}% bold math
\usepackage{amsmath,amssymb}
\usepackage[dvips]{epsfig}
\begin{document}

%%%%%%%%%%%%%%%%%%%%
\title{Directed current in the Holstein system}
\author{D. Hennig}
\author{A.D. Burbanks}
\author{A.H. Osbaldestin}
\affiliation
{Department of Mathematics, University of Portsmouth, Portsmouth, PO1 3HF, UK}

%\pacs{40$-$a}

\begin{abstract} We propose a mechanism to rectify charge transport in the semiclassical Holstein model.  It is shown that localised initial conditions, associated with a polaron solution, in conjunction with a nonreversion symmetric static electron on-site potential 
constitute  minimal prerequisites for the emergence of a directed current in the underlying periodic lattice system. In particular, we demonstrate that for unbiased spatially localised initial conditions, violation of parity prevents the existence of pairs of counter-propagating trajectories, thus allowing for a directed current despite the time-reversibility of the equations of motion.
Occurrence of long-range coherent charge transport is demonstrated.
\end{abstract}

\maketitle

\date{\today}

%\underline{The Holstein model with an asymmetric potential}

Since the pioneering work of Landau and Pekar it is established
that when an electron in a lattice interacts locally with the
phonons it can become self-trapped by the lattice distortion it
creates \cite{Landau},\cite{Pekar}. The generated quasi-particle
consisting of the localised electron and its associated local
lattice deformation is called a polaron (an electron surrounded by
a phonon cloud). Fr\"ohlich (large polarons) and Holstein (small
polarons) elaborated significantly on the consequences of the
polaron concept \cite{Froehlich},\cite{Holstein}.  Later on this idea of a
localised quasi-particle was used by Davydov to propose a model
for coherent energy and/or charge transport in macromolecules
\cite{Davydov}-\cite{Chetverikov}.

Recently the emergence of a directed current triggered by an external time-dependent field with zero mean which can be of stochastic or deterministic nature has attracted considerable interest. Applications include charge transport
in semiconductor heterostructures~\cite{hetero}, current generation in semiconductor
superlattices~\cite{super}, nano-engines~\cite{nano}, directed collective energy current in spatially extended systems \cite{extended},
superconductors~\cite{supercon}, spin transport~\cite{spin}, and the
motion of cold atoms in driven optical potentials~\cite{optics}.

The necessary conditions for rectification of the current, based on
symmetry investigations of the external field and the underlying
static potential, have been presented in~\cite{super}, \cite{Flach},
and~\cite{Denisov}. To be precise, all symmetries that, to each
trajectory, generate a counterpart moving in the opposite direction,
need to be broken.  Furthermore, the phase space has to be mixed,
with coexisting regular and chaotic
dynamics~\cite{Schanz}.  This is achievable by imposing a
time-dependent external force that is periodic but not symmetric under
time reversal~\cite{Flach}--\cite{Schanz1}.  In extended chaotic
systems a nonzero current can be obtained as the time-averaged
velocity of an ensemble of trajectories in the chaotic component of
phase space and the chaotic transport proceeds ballistically and
directedly~\cite{Schanz}, \cite{Schanz1}. 
Recently the emergence of directed flow in autonomous Hamiltonian systems has been demonstrated in \cite{Hennig}. There it is shown that internal parity violation and  localised initial conditions, which are unbiased,  in conjunction with transient chaos conspire to form the physical mechanism for the occurrence of a current.

In this manuscript we propose a symmetry-breaking mechanism to accomplish directed electron (charge) transport in the context of the semiclassical Holstein system. 
The Hamiltonian of the semiclassical Holstein system consists of three parts
\begin{equation}
H=H_{el}+H_{vib}+H_{int}\,.
\end{equation}
The first term
\begin{equation}
H_{el}=\sum_{n}\,\left\{ U_n |c_n|^2- V \left(\,c_{n}^{*} c_{n-1}+c_n
c_{n-1}^{*}\,\right)\right\} \,.\label{eq:Hel}
\end{equation}
describes quantum-mechanically the motion of an excess electron in the tight-binding approximation over the units of a one-dimensional chain where
$c_n$ determines the
probability amplitude to find the electron residing at site $n$.
$V$ is the transfer matrix element (its value is
determined by an overlap integral) being responsible for the
nearest-neighbour transport of the electron along the chain and $U_n$ denotes the on-site potential (on-site energy of the electron).
The second term $H_{vib}$ represents the Hamiltonian for the classical dynamics of
(local) Einstein oscillators
\begin{equation}
H_{vib}= \sum_{n}\,\left\{\,\frac{p_{n}^{\,2}}{2m}
\,+\,\frac{m\omega_0^2}{2} q_n^{\,2}\,\right\}\,.\label{eq:Hvib}
\end{equation}
The coordinates $q_{n}$ quantify the displacements from equilibrium position of the oscillator at site $n$, $p_n$ is the corresponding canonically conjugate momentum, $\omega_0$ is the harmonic vibrational frequency, and $m$ is the mass of an oscillator. The interaction between the electronic and the vibrational degrees
of freedom is due to the modifications of the electronic
on-site potential  $U_{n}$ by displacements of the oscillators from
their equilibrium positions and is is described by the
following interaction Hamiltonian
\begin{equation}
 H_{int}=\alpha\sum_{n}\,q_n |c_n|^2\,,\label{eq:Hint}
\end{equation}
where $\alpha$ regulates the electron-oscillator coupling strength.

For a dimensionless representation we introduce the following time
scale:
${\tilde t} = \omega_0\,t$.
The dimensionless representation of the remaining variables and the  parameters of the
system follows from the relations:
\begin{equation}
\tilde{q}_{n}= \sqrt{\frac{m\omega_0^2}{V}} \, q_{n}\,,
\qquad \tilde{\alpha}=\frac{\alpha}{\sqrt{ m \omega_0^2 V}}\,,\qquad \tilde{U}_n=\frac{U_n}{V}\,.
\end{equation}
In what follows we drop the tildes.

The equations of motion derived from the Hamiltonian given in Eqs.
(\ref{eq:Hel})--(\ref{eq:Hint}) read as
\begin{eqnarray}
i\,\tau \frac{dc_n}{dt} &=&
U_n c_n  - (c_{n+1}+ c_{n-1})+\alpha q_n c_n \label{eq:cq}\\
\frac{d^2q_{n}}{dt^2}\,&=&\,-q_{n}-\alpha |c_n|^2 \,.\label{eq:qc}
\end{eqnarray}
The  parameter $\tau=\hbar \,\omega_0/V$,
appearing in the l.h.s. of Eq.\,(\ref{eq:cq}), determines the
degree of time-scale separation between the (slow) electronic and
(fast) vibrational processes.

\vspace*{0.5cm}
%\underline{Polaron solutions}

In the following we construct standing localised solutions. To this end we consider the corresponding stationary system. For the oscillator system the condition $\ddot{q}_n=0$ yields the stationary displacements
\begin{equation}
 \hat{q}_n=-\alpha |c_n|^2\label{eq:instant}
\end{equation}
from which, after substitution
in Eq.\,(\ref{eq:cq}), one obtains the following discrete nonlinear Schr\"odinger equation
\begin{equation}
i\,\tau \frac{dc_n}{dt} =
U_n c_n  - (c_{n+1}+ c_{n-1})-\gamma |c_n|^2 c_n\,, \label{eq:dnls}
\end{equation}
with $\gamma=\alpha^2$.
Substituting the ansatz $c_n(t)=\phi_n \exp(-i E/\tau t)$ with $\phi_n \in \mathbb{R}$ in Eq.\,(\ref{eq:dnls}) one arrives at the system of coupled difference equations
\begin{equation}
 E\phi_n=U_n \phi_n-(\phi_{n+1}+\phi_{n-1})-\gamma \phi_n^3\,.\label{eq:difference}
\end{equation}

\begin{figure}
\begin{center}
\includegraphics[height=6cm, width=8cm]{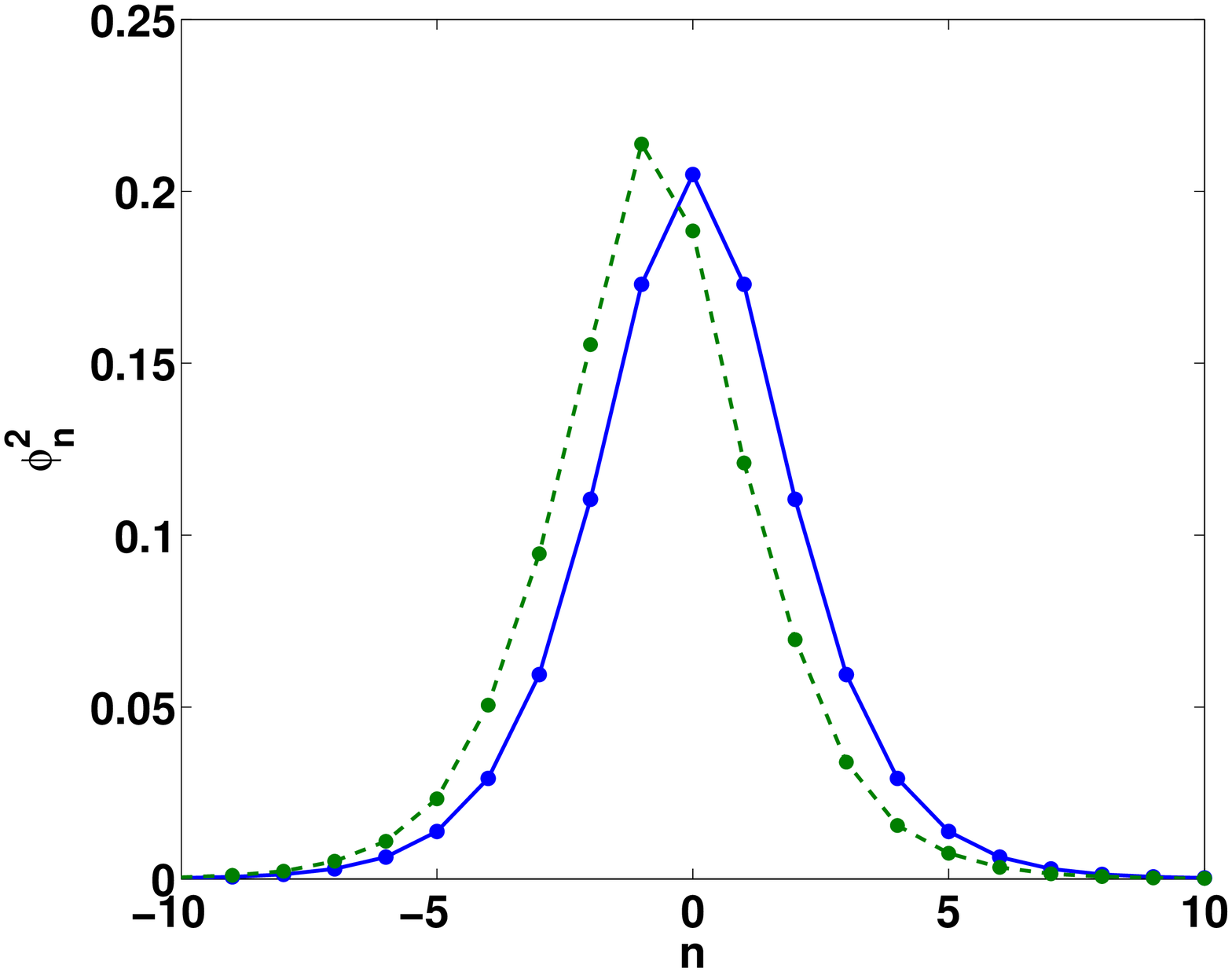}
\end{center}
\caption{Profile of the stationary electron occupation probability displayed for a lattice segment around the central site $n_0=0$. The parameter values are given by  $\alpha=1.25$ and $\omega_0^2=2$. Solid line: The symmetrical on-site potential with $a=U_0=0$, Dashed line: 
The asymmetrical on-site potential with $a=0$ and $U_0=0.035$.} \label{fig:profile}
\end{figure}

This system supports localised solutions corresponding to standing localised electrons.
Following the methods outlined in Refs. \cite{Kalosakas},\cite{Voulgarakis} we derive the profile of the standing localised electron wave function. 
Subsequently, with the help of Eq.\,(\ref{eq:instant}) one obtains the corresponding localised lattice oscillators forming in conjunction with the localised electron the polaron compound. One utilises that polaron solutions are obtained as the attractors of the map
\begin{equation}
\{\phi\} \rightarrow \{ \bar{\phi} \}=H\{\phi\}/\lVert H\{\phi\}\rVert\,,
\end{equation}
where the operator $H$ is determined by the right hand side of Eq.\,(\ref{eq:difference}) and the norm of the state $H\{\phi\}$ is defined as $\lVert H\{\phi\}\rVert=\sqrt{\sum_n(H\{\phi\})^2}$. The map iterations are started with a completely localised state, i.e. $\{\phi_n^{(0)}\}=\delta_{n,n_0}$, and act on it with the operator $H$. After each application of $H$ the resulting vector is normalised and the iteration procedure is terminated when convergence is attained yielding the polaron state of lowest energy.
In Fig.~\ref{fig:profile}
we show
the profile of the electron occupation probability $\phi_n^2$ for two different
situations. The first case is for uniform electron on-site potential  $U_n=0$ (solid line in Fig.~\ref{fig:profile}).
(For $U_n=U={\rm constant}$, a simple gauge transformation
$c_n(t)=\bar{c}_n(t)\exp(iU t)$ removes the on-site term
from the equation of motion.)  
The peak of the profile is located at the central site $n_0=0$ and the occupation probability decays to either side in an exponential fashion.
Note the inversion symmetry with respect to this site, i.e. $\phi_{n}=\phi_{-n}$. For the second case we chose a sawtooth-like profile of the electron on-site potential of the following form
\begin{equation}
U_n-a=\begin{cases} U_0, & n=3k;\\[2mm]
2U_0, & n=3k+1;\\[2mm]
3U_0, & n=3k+2\,,\\
\end{cases}\label{eq:ratchetpotential}
\end{equation}
where $k=0,\pm 1,\pm 2,...$. Notice that an electronic on-site potential of the form in (\ref{eq:ratchetpotential}) breaks the reflection symmetry $n_0+n \mapsto n_0-n$ of the lattice system  inducing in this way an asymmetric potential biasing motion to the right. The lattice system is still periodic exhibiting translational invariance. 
In fact this potential possesses the shortest possible 'unit cell', namely of period-$3$, violating the reflection symmetry of the lattice system.
The resulting asymmetry is reflected in the profile of the static polaron solution which for  an asymmetric potential with $a=0$ and $U_0=0.035$ is depicted by the dashed line in Fig.~\ref{fig:profile}. 
Remarkably the peak of the occupation probability is not situated at the central site $n_0=0$ but is shifted one lattice site to the left. As in the symmetrical case the occupation probability is exponentially localised at a single site.

For later use we introduce here for a lattice comprising $N$ sites the participation number, $P$, defined as
\begin{equation}
 P=\frac{1}{\sum_{n=1}^{N} \phi_n^4}\,,
\end{equation}
quantifying the spatial extension of the electron occupation probability (degree of electron localisation). An electron completely localised at a single lattice site corresponds to $P=1$ whereas $P=N$ corresponds to a completely delocalised electron for a finite lattice of $N$ sites. For the occupation probabilities displayed in Fig.~\ref{fig:profile} one obtains $P\simeq 7$ for both the symmetrical and the asymmetrical case.

For a quantitative assessment of directed transport we consider the
current density $J$ being proportional to
\begin{eqnarray}
 J(t)&=&\frac{1}{N}\sum_{n=1}^N\,J_n(t)\nonumber\\
&=&i\frac{1}{N}\,\sum_{n=1}^N\,{\rm{Im}}
\left[c_n^{*}(t)\left(c_{n+1}(t)-c_{n-1}(t)\right)\right]\,.
\end{eqnarray}
In the following we discuss the symmetry features of the lattice system. We are in particular interested in the emergence of a directed current facilitated by the asymmetric potential.

\vspace*{0.5cm}
%\underline{Symmetries}

In order to obtain a non-zero current (net flow), all spatiotemporal symmetry operations that leave the system of equations of motion invariant and invert the sign of the current density must be broken. Otherwise the current density $J$ completely vanishes.
Regarding the symmetries of the semi-classical equations of motion  we note that Eqs.\,(\ref{eq:cq}),(\ref{eq:qc}) are invariant under the time-reversal operation $t \mapsto -t$.  For the canonically conjugate momentum and coordinate variables of the classical system of Einstein oscillators this is equivalent to the transformation $(\{p_n(t)\},\{q_n(t)\}) \mapsto (\{-p_n(-t)\},\{q_n(-t)\})$. For the
quantum discrete Schr\"odinger equation the time-reversal operation, $t\mapsto -t$, is connected with complex conjugation of the probability amplitudes, $c_n$, which also reverses the sign of the current density. 
Passing from the pairs of complex-valued canonically conjugate variables, 
$\{c_n, c_n^*\}$, to  real-valued canonically conjugate momentum and coordinate variables via the canonical transformation $c_n=(Q_n+i\,P_n)/\sqrt{2}$ 
and using the representation $c_n=r_n\exp(i\theta_n)$ one obtains the relations
$P_n=\sqrt{2}r_n\sin(i\theta_n)$ and $Q_n=\sqrt{2}r_n\cos(i\theta_n)$. Upon time-reversal operation, $t \mapsto -t$, the corresponding inversion of the sign of the phases after complex conjugation, i.e. $c_n^*=r_n\exp(-i\theta_n)$, is  equivalent to the transformation $(\{P_n(t)\},\{Q_n(t)\}) \mapsto (\{-P_n(-t)\},\{Q_n(-t)\})$\,.

In general, solutions of the system are of the form 
\begin{equation}
X(t)=(\{p_n(t)\},\{q_n(t)\},\{c_n^*(t)\},\{c_n(t)\})\,.
\end{equation}
Thus applying the time-reversal operator yields 
\begin{eqnarray}
& &\hat{\tau}(\{p_n(t)\},\{q_n(t)\},\{c_n^*(t)\},\{c_n(t)\})\nonumber\\
&=&\{(-p_n(-t)\},\{q_n(-t)\},\{c_n(-t)\},\{c_n^*(-t)\}),
\end{eqnarray} 
and hence, if $X$ is a solution, then so is $\hat{\tau}X$.
As for the implication of
time-reversibility symmetry with regard to the net flow, let a
solution, starting from some initial condition ${X}(0)$, be evolved in
time up to a finite observation time $T$ at which the {\it forward}
trajectory arrives at the point ${X}(T)$ on the constant energy
surface. Subsequently, under application of the time-reversal
operation, for the classical subsystem the signs of the momenta $\{p_n\}$ at this point, ${X}(T)$, are
reversed while the complex conjugation is applied to the probability amplitudes $\{c_n\}$. Letting then the solution evolve once again, with
$\widehat{\tau}X(T)$ as the initial condition, the corresponding {\it
  backward} trajectory traces back the path of the forward trajectory
in coordinate space $(\{q_n\},\{Q_n\}={\rm Re}(\{c_n\}))$. 
For a microcanonical ensemble, the initial
conditions ${X}(0)$ and $\widehat{\tau}{X}(T)$ are equally selected
points from the constant energy surface.  Thus, for systems with
time-reversibility symmetry and {\it uniformly distributed initial
conditions} populating the whole energy surface there is no preferred
direction of the flow thus preventing the emergence of a current.

However, considering a static polaron solution as the initial state for the dynamics
corresponds to initial conditions being exponentially localised at a specific site, $n_0$, of the lattice. Therefore, only a confined region of the energy surface contains initial conditions. Suppose the directed motion of a polaron can be instigated; then
the polaron undergoing directed motion leaves the domain of {\it localised initial conditions} on the lattice
(the extension of which is determined by the participation number, $P$, defined above) and contributes to a net current. Crucially, at the end of the observation time, $T$, the polaron (assuming it travels directedly far enough while retaining its localised shape) lies outside the domain of the localised initial conditions. Consequently, the initial condition of the associated backward moving polaron, compensating the contribution of the forward moving polaron to the net current, is {\it not contained} in the set of localised initial conditions.  We underline
 that this alone does not imply the emergence of a current in
the system for such sets of initial conditions \cite{Hennig}. 

In fact, in order to obtain a non-zero current density the remaining invariances of the system, which induce  symmetries ensuring that there exist trajectories which mutually compensate each others contribution to the net flow, must be broken.

In this context we note that the system exhibits reflection symmetry $n_0+n \mapsto n_0-n$  if and only if the electronic on-site potential $U_n$ possesses reflection symmetry. In more detail, the following parity symmetry operations leave the system of equations of motion invariant and reverse at the same time the sign of $J$:
\begin{eqnarray}
n_0+n \mapsto n_0-n\,,& &\qquad U_{n_0+n}\mapsto U_{n_0-n}\,,\\
c_{n_0+n} \mapsto c_{n_0-n}\,,& &\qquad q_{n_0+n}\mapsto q_{n_0-n}\,.
\end{eqnarray}
As a consequence, in the presence of parity (reflection) symmetry there exist pairs of counterpropagating trajectories,
\begin{equation}
 X_{+}(t)=(\{p_{n_0+n}(t)\},\{q_{n_0+n}(t)\},\{c_{n_0+n}(t)\},\{c_{n_0+n}^*(t)\})
\end{equation} 
and 
\begin{equation}
X_{-}(t)=(\{-p_{n_0-n}(t)\},\{q_{n_0-n}(t)\},\{c_{n_0-n}^*(t)\},\{c_{n_0-n}(t)\}) 
\end{equation}
annihilating each others contribution to the net current density.

Crucially, an asymmetrical electron potential such as the one of the form in (\ref{eq:ratchetpotential}) breaks the parity symmetry and thus allows for the occurrence of a non-zero current density.

\vspace*{0.5cm}
%\underline{Mobile polarons and current emergence}

So far we dealt with standing polaron solutions having zero associated kinetic energy.
As a first step we initiate the motion of the 
polaron through suitable initial conditions of the momenta of the lattice oscillators  $\{p_n(0)\}$ targeted in the direction of  the pinning mode \cite{Chen}. 
To be precise we use the following initial conditions for the numerical integration of the system;
\begin{equation}
\left\{q_n(0),p_n(0),{\rm Re}(c_n(0)),{\rm Im}(c_n(0))\right\}=
\left\{\hat{q}_n,\lambda \xi,\phi_n,0\right\}\,, 
\end{equation}
with the normalised momentum part $\xi$ of the pinning mode and $\lambda$ regulating its amplitude. Here, we exploit the lowest frequency pinning mode although others of higher frequency could also be applied.
We solve numerically the set of coupled equations using a Runge-Kutta method for a lattice consisting of $N=105$ sites. The norm conservation $\sum_n |c_n(t)|^2=1$ (as well as the conservation of energy) was monitored during the integration procedure to guarantee accurate computations. 
We employ periodic boundary conditions. We stress the relation between  the extension of the polaron and the length of the lattice $P \ll N$.

\begin{figure}
\includegraphics[height=7cm, width=8cm]{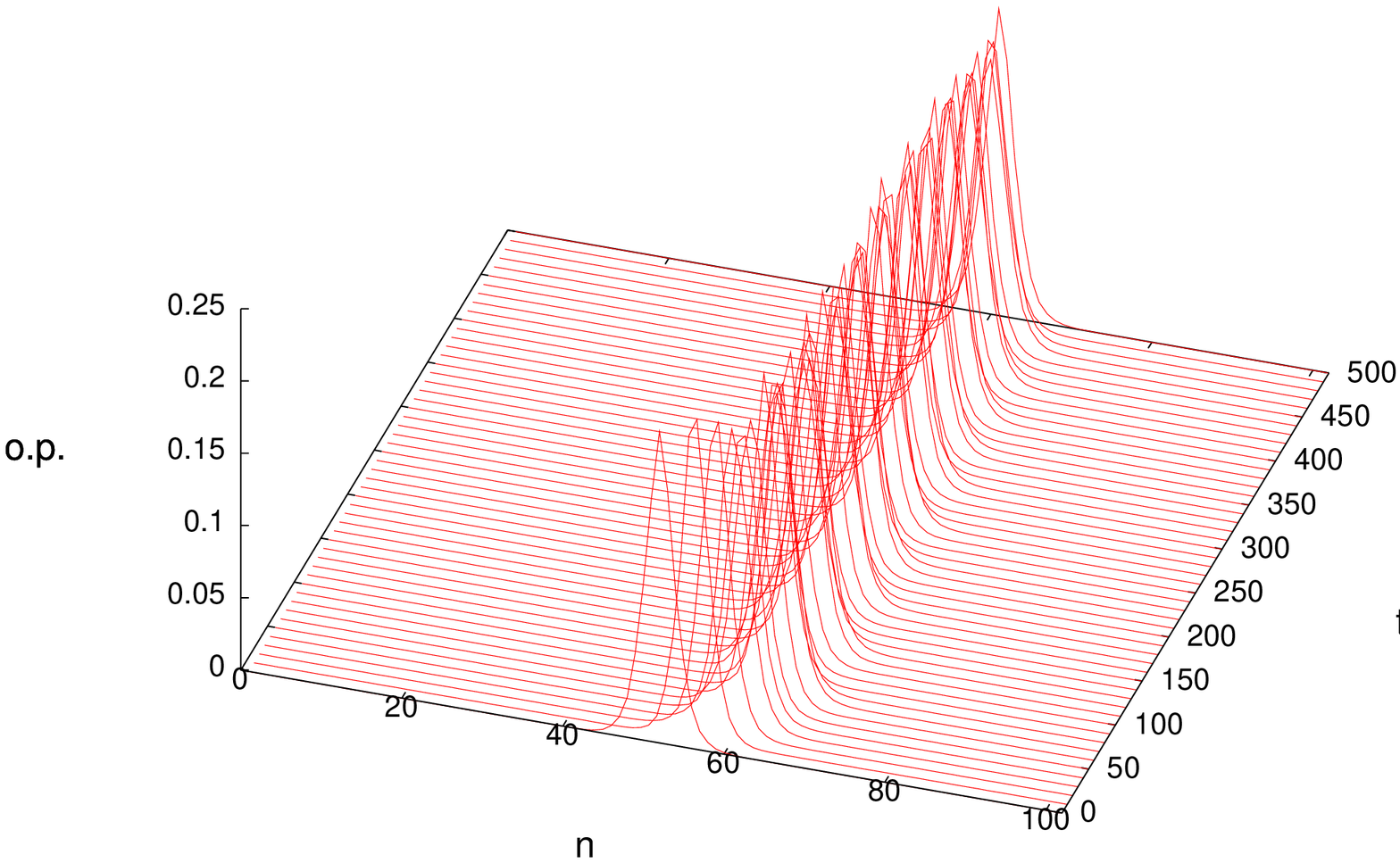}
\includegraphics[height=7cm, width=8cm]{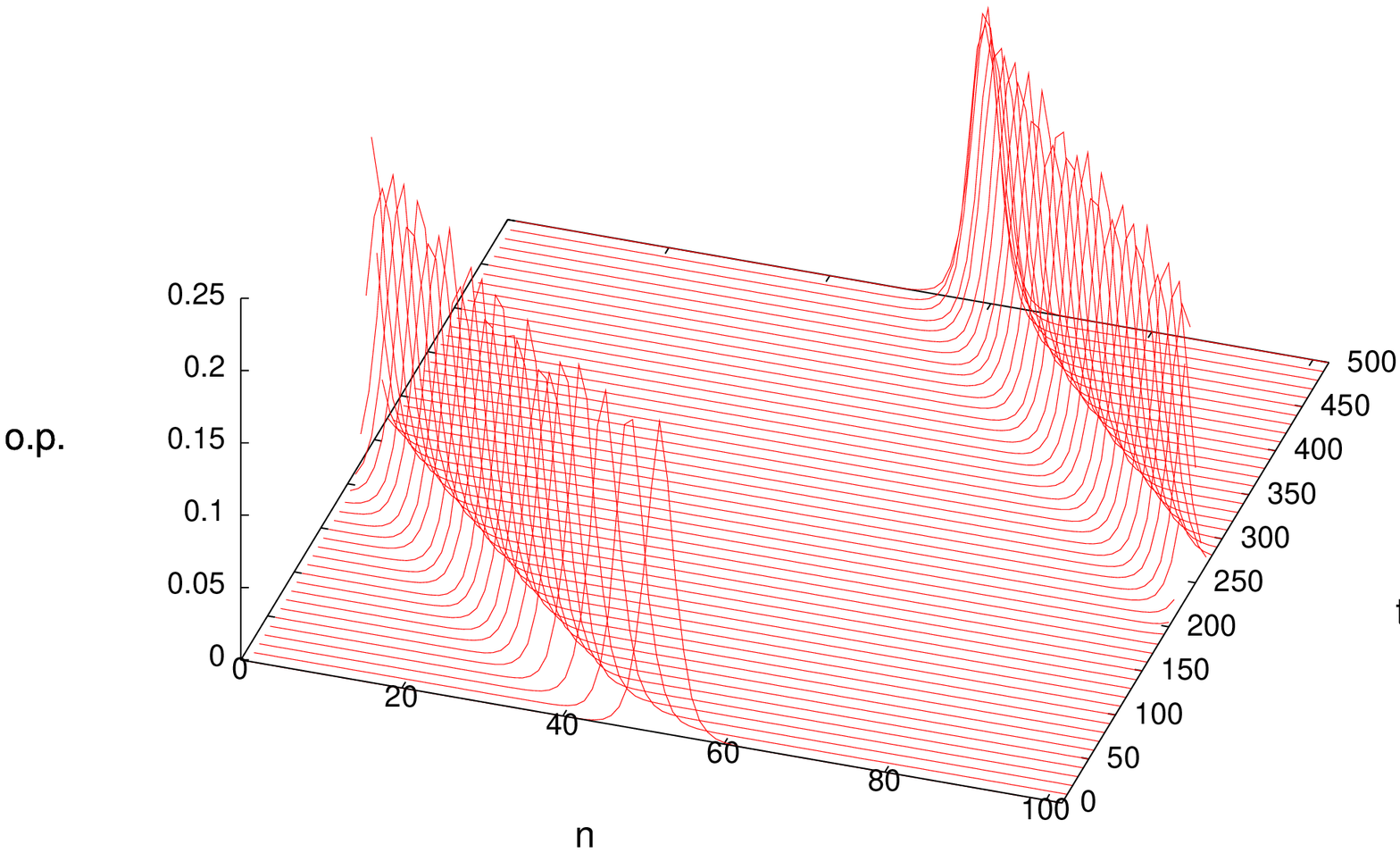}
\caption{Spatiotemporal evolution  of the electron occupation probability (o.p.) for a lattice comprising $N=105$ sites and periodic boundary conditions are imposed. The parameter values are given by $\alpha=1.25$, $\omega_0^2=2$, $\tau=25$, $a=0$, and $U_0=0.035$. Upper (lower) panel $\lambda >0$ ($\lambda <0$).} \label{fig:evolution}
\end{figure}

The effect of asymmetry becomes apparent in Fig.~\ref{fig:evolution} which displays the spatiotemporal evolution of the electron occupation probability for pinning mode amplitudes $\lambda=\pm \sqrt{0.15}$. We underline that interchanging the sign of the amplitude $\lambda \leftrightarrow -\lambda$ entails merely the reflection anti-symmetry $p_n(0) \leftrightarrow -p_n(0)$ in the initial momenta under preservation of the kinetic energy, thus, imposing {\it no bias}.
For the pattern shown in the top panel (positive $\lambda$)  the electron tends to move to the right in an initial phase. However, it soon becomes trapped at a lattice site. In stark contrast with this, its counterpart in the bottom panel (negative $\lambda$) exhibits coherent motion to the right under maintenance of the localised pattern (despite the initial slight broadening of the profile). Strikingly, radiation losses due to the emission of phonons seem to be suppressed.
In this case the electron is accordingly accompanied by the associated localised pattern of lattice oscillators (not shown).

\begin{figure}
\begin{center}
\includegraphics[height=6cm, width=8cm]{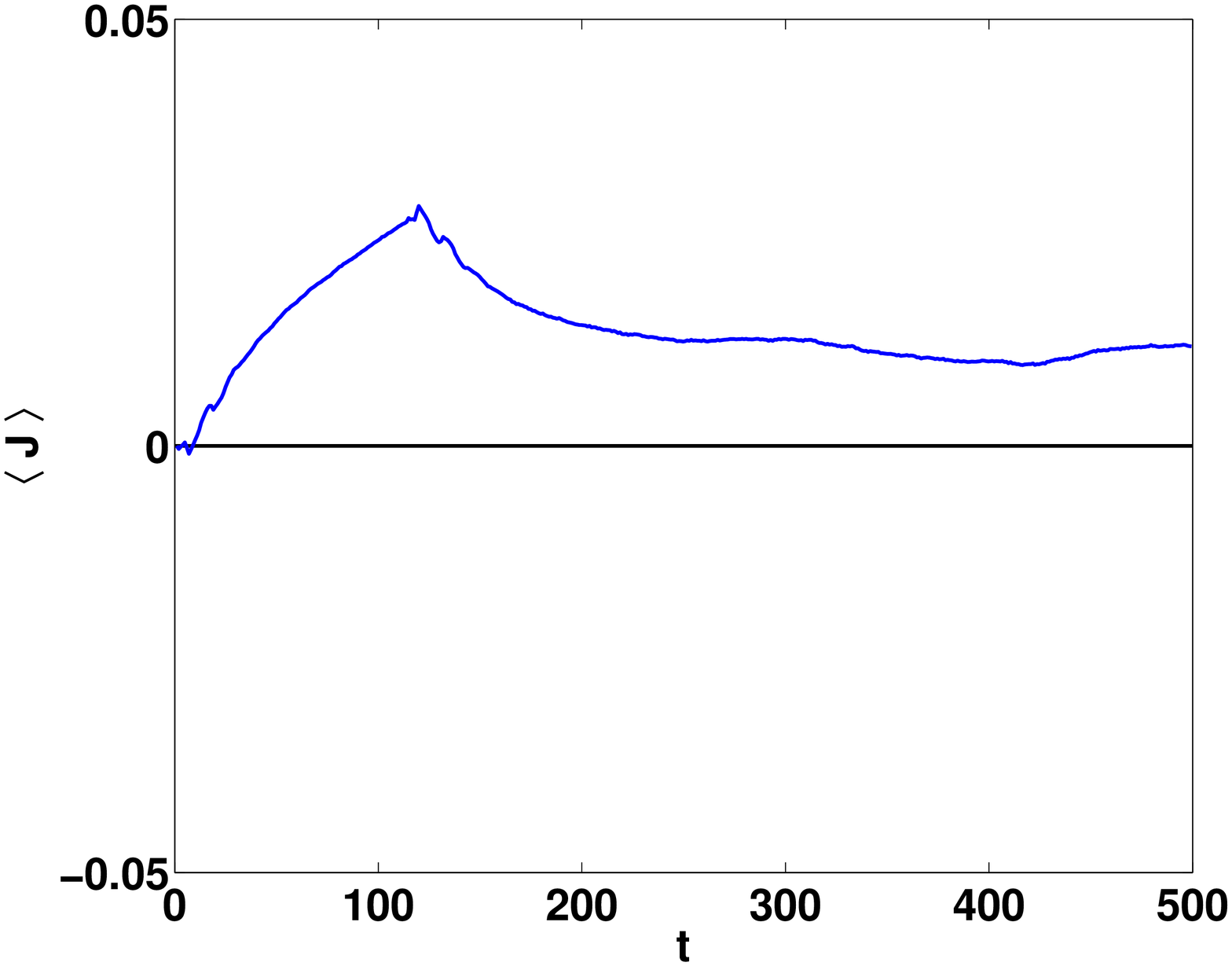}
\end{center}
\caption{Temporal behaviour of the mean current density $\langle J(t)\rangle$ for an ensemble of $1000$ initial conditions distributed according to Eq.\,(\ref{eq:initial}) with $E_{kin}=0.1$. 
The parameter values are given by $\alpha=1.25$, $\omega_0^2=2$, $\tau=25$, $a=0$, and $U_0=0.035$.} \label{fig:current}
\end{figure}

Furthermore, we utilised ensembles of {\it unbiased} initial conditions for the momenta to  instigate the motion of polarons.  To be precise, kinetic energy is provided locally to the system
as follows: In the region of the lattice supporting the localised standing-electron lattice-oscillator compound whose spatial extension is determined by the participation number $P$, kinetic energy of magnitude $E_{kin}$ is injected in  such a way that the initial values of the momenta, $\{p_n(0)\}$,  are distributed uniformly on an iso-energetic sphere in momentum-space such that
the relation
\begin{equation}
E_{kin}=\frac{1}{2}\sum_{n=-[P/2]}^{[P/2]} p_{n_0+n}^2\,.\label{eq:initial}
\end{equation}
is fulfilled and $[\,\cdot\,]$ denotes the integer part of $P/2$.
Notice the symmetry $p_i\leftrightarrow -p_i$ and $i=n_0-[P],...,n_0+[P]$. Hence there is {\it no bias} contained in the ensemble of initial conditions $\{ p_n(0) \}$. We emphasise however the localised character of the initial conditions.
For a quantitative assessment of the efficiency of the transport-rectifying mechanism we consider the
mean current density $\langle J \rangle$ where $\langle \cdot \rangle$ denotes the ensemble average. 
In Fig.~\ref{fig:current} we show the temporal behaviour of the mean current density for our simulation for which the motion of the standing asymmetrical polaron shown in Fig.~\ref{fig:profile} is instigated with the inclusion of kinetic energy in the initial conditions in the lattice segment $n\in [-4,4]$. To this end an ensemble of $1000$ initial conditions with $\{p_n(0)\}$ fulfilling the relation  in Eq.\,(\ref{eq:initial}) with $E_{kin}=0.1$ is used. Note that the amount of  kinetic energy injected is small compared to the energy of the asymmetrical polaron, which amounts to $E_{polaron}=\sum_n[U_n \phi_n^2-2\phi_n\phi_{n-1}-(\gamma/2)\phi_n^4]\simeq -1.965$.
Notably,
after an initial transient  where the current density grows continually towards larger positive values, it reaches an almost stationary regime characterised by subdued variations. Most importantly, $J$ 
stays positive throughout the simulation time indicating a positive net current density complying with the bias direction induced by the asymmetric potential. On the other hand, for restored reflection symmetry, i.e. $U_{n_0+n} \mapsto U_{n_0-n}$, there results zero net current density.

In summary, we have proposed a transport-rectifying mechanism accomplishing directed charge transport in the semiclassical Holstein model. We have shown that using localised initial conditions, constituted by a polaron solution, together with a nonreversion symmetric electron on-site potential suffices for the rectification of the current. Crucially, violation of parity prevents the existence of pairs of counterpropagating trajectories emanating from  an ensemble of unbiased localised initial conditions despite the time-reversibility of the underlying equations of motion. Our numerical simulations of the semiclassical Holstein system with an asymmetric on-site potential reveal a regime of coherent long-range charge transport. Our method can directly be applied to induce directed transport in other lattice systems supporting (mobile) localised solutions.

%\newpage

\end{document}